\documentclass[prd,twocolumn,nofootinbib,reprint]{revtex4-1}
\usepackage{amsmath,amssymb,amsthm,bm,braket,ascmac,bbm}
\usepackage{graphicx}

\theoremstyle{definition}

\begin{document}
\title{
A New Scheme for NMSSM in Gauge Mediation
}
\author{Masaki Asano$^{1}$, Yuichiro Nakai$^{2}$, Norimi Yokozaki$^{3}$}
\address{
\vspace{12pt}$^{1}$Physikalisches Institut and Bethe Center for Theoretical Physics, Universit\"at Bonn, Nussallee 12, D-53115 Bonn, Germany \\
$^{2}$Department of Physics, Harvard University, Cambridge, Massachusetts 02138, USA\\
$^{3}$Istituto Nazionale di Fisica Nucleare, Sezione di Roma, Piazzale Aldo Moro 2, I-00185 Rome, Italy \vspace{5pt}}

\begin{abstract}
\begin{center}{\bf Abstract}\end{center}

We propose a new framework for the next-to-minimal supersymmetric standard model (NMSSM) in gauge mediation,
where in general the correct electroweak symmetry breaking (EWSB) is difficult to be explained. 
The difficulty is caused by the absence of a soft supersymmetry (SUSY) breaking mass for the NMSSM singlet $S$.
In our framework, $S$ is a meson in a hidden QCD. This QCD is responsible for the dynamical SUSY breaking, 
forming $S$, and the soft SUSY breaking mass for $S$, which is a key to explain the correct EWSB: all the ingredients for successful phenomenology originate from the common dynamics. 
From the requirement of the successful EWSB, the low-scale SUSY breaking around 100-1000\,TeV is predicted. 
This is favored to avoid the large fine-tuning. 



\end{abstract}
\pacs{***}
\maketitle

\section{Introduction}\label{sec:intro}

Supersymmetry (SUSY) is a promising candidate beyond the standard model (SM). Dynamical SUSY breaking gives a beautiful explanation for the question why the electroweak scale is so small compared to the Planck scale or the unification scale \cite{Witten:1981nf} (For a review, see \cite{Intriligator:2007cp}) and the gauge mediation mechanism~\cite{gmsb,gmsb_old} 
(For reviews, see e.g. \cite{Giudice:1998bp,Kitano:2010fa}) can naturally explain why dangerous flavor-changing neutral currents are highly suppressed.\footnote{
In this sense, SUSY models explaining the muon $g-2$ anomaly based on gauge mediation~\cite{gmsb_gm2} 
are more convincing than those based on gravity mediation. 
This is because the light slepton and chargino/neutralino are always required.
} 

However, the observed Higgs mass, 125 GeV, may conflict to a minimal realization of such a  scenario \cite{Draper:2011aa}. To obtain the observed Higgs mass in the minimal supersymmetric standard model (MSSM), a significant radiative correction from top/stop loops \cite{Haber:1990aw,Okada:1990vk,Barbieri:1990ja} are required, then, in the gauge mediation scenario, sparticle masses should be very large because there is no large stop trilinear coupling.

The simplest modification would be to add a SM singlet chiral superfield $S$. In such a model, the $S H_u H_d$ superpotential interaction provides an additional $F$-term contribution to the Higgs potential and it can push the lightest Higgs boson mass up without large stop contributions. This is known as the next-to-minimal supersymmetric standard model (NMSSM) and have been widely investigated so far (For reviews, see e.g. \cite{Maniatis:2009re,Ellwanger:2009dp}). 

Although the additional $F$-term contribution to the Higgs potential is appealing, the viable parameter space is restricted because large singlet couplings, e.g., $S H_u H_d$ tends to blow up at high energy. Furthermore, such a singlet extension of the gauge mediation scenario is not straightforward: since the singlet does not have any SM gauge charge, 
the SUSY breaking hardly mediates to the singlet sector, which makes it difficult to achieve the correct electroweak symmetry breaking (EWSB); therefore, further extensions, e.g., introducing extra vector-like matter fields coupling to $S$~\cite{murayama_nmssm, nmssm_domain},\footnote{
In Refs.~\cite{nmssm_domain}, it has been pointed out that the domain wall problem in the $Z_3$ invariant NMSSM is solved with these new vector-like matter fields, which make $Z_3$ anomalous.
%
} and the coupling between $S$ 
and messengers~\cite{nmssm_gmsb_xis, giudice_slavich, nmssm_gmsb_higgs} have been studied.\footnote{
See also Ref.\,\cite{Kowalska:2015wua}.
}


The solution may be ``hidden". In this paper, we propose a possibility that the NMSSM singlet is a meson in the hidden sector. This meson is composed by particles which are charged under a strong gauge symmetry as in Fat Higgs models~\cite{Harnik:2003rs,Chang:2004db}. The dynamics of the strong gauge symmetry provides a meta-stable SUSY breaking vacuum around the origin of the field space \cite{Intriligator:2006dd} and messengers of gauge mediation also couple to the SUSY breaking sector \cite{Murayama:2006yf}. The $S H_u H_d$ coupling is provided by integrating out some heavy particles charged under the strong gauge symmetry, then, the particle decoupling is also a trigger of the confining dynamics and creates a meta-stable SUSY breaking vacuum. 

Since the singlet is also a member of the hidden sector, the SUSY breaking can mediate to the singlet directly. In fact, the singlet receives the SUSY breaking at two loop and it plays an important role to achieve the correct EWSB in our scenario. Although this is not a gauge-mediated SUSY breaking, such a soft mass of the singlet does not cause dangerous flavor-changing neutral currents either. 

Supposing the TeV superparticle mass spectrum, a viable SUSY breaking scale will be about $\mathcal{O}(100)$-$\mathcal{O}(1000)$ TeV in our scenario. This would be also an appealing prediction. Such a $\mathcal{O}(100)$ TeV SUSY breaking provides a cosmologically safe gravitino mass of order eV. A large $S H_u H_d$ coupling is possible because the singlet composite scale is not very far from the electroweak scale. 
Also by the naturalness discussion of the EWSB~\cite{ns_asano}, such a low messenger scale is favored because radiative corrections to the Higgs potential can be relatively small due to the short running. 

The rest of the paper is organized as follows. In the next section, we present our model and discuss the formation of the NMSSM singlet and the dynamical SUSY breaking in that model. The mediation of the SUSY breaking to ordinary superpartners is also discussed. In section III, we show expected mass spectra at the electroweak scale and investigate the phenomenology of the model. In section IV, we conclude the discussion and comment on possible future directions.

\section{The framework}\label{sec:model}

In this section, we first present our model and describe
the formation of the NMSSM singlet and dynamical SUSY breaking at a meta-stable vacuum.
A 2-loop radiative correction from the SUSY breaking sector
generates a negative soft mass-squared for the singlet.
We also explain gauge-mediated SUSY breaking for ordinary superparticles.

\subsection{Composite NMSSM}

\renewcommand{\arraystretch}{1.3}
\begin{table}[!t]
\begin{center}
\begin{tabular}{c|cccc}
 & $SU(N)_{H}$ & $SU(3)_{C}$ & $SU(2)_{L}$ & $U(1)_Y$ 
 \\
 \hline
  $Q_I$ ($I = 1, \cdots , N_f$)&  $\mathbf  N$ & $\mathbf 1$ & $\mathbf  1$ & $0$ \\
  $\bar{Q}_I$ ($I = 1, \cdots , N_f$)& $\mathbf{\bar{N}}$ & $\mathbf 1$ & $\mathbf  1$ & $0$ \\
  ${\Phi}_c$ & $\mathbf{{1}}$ & $\mathbf 3$ & $\mathbf  1$ & $-1/3$ \\
  $\bar{\Phi}_c$  &  $\mathbf  1$ & $\mathbf{\bar{3}}$ & $\mathbf  1$ & $1/3$ \\
  ${\Phi}_l$ & $\mathbf{ {1}}$ & $\mathbf 1$ & $\mathbf  2$ & $1/2$ \\
  $\bar{\Phi}_l$  &  $\mathbf  1$ & $\mathbf 1$ & $\mathbf  2$ & $-1/2$ \\
    $f$ & $\mathbf{ {N}}$ & $\mathbf 3$ & $\mathbf  1$ & $-1/3$ \\
  $\bar{f}$  &  $\mathbf{\bar N}$ & $\mathbf{\bar 3}$ & $\mathbf  1$ & $1/3$ \\
    $\Psi_u$  &  $\mathbf  N$ & $\mathbf 1$ & $\mathbf  2$ & $1/2$ \\
  $\bar{\Psi}_d$ & $\mathbf{\bar{N}}$ & $\mathbf 1$ & $\mathbf  2$ & $-1/2$ \\
  $X_m$, $Y_m$  &  $\mathbf 1$ & $\mathbf 1$ & $\mathbf  1$ & $0$ 
\end{tabular}
\end{center}
\caption{The matter content and charge assignment of a model of the composite NMSSM with dynamical SUSY breaking
and its gauge mediation.}
\label{tab:model}
\end{table}
\renewcommand{\arraystretch}{1}

Let us consider a supersymmetric $SU(N)_H$ gauge theory with $(N + 6)$ vector-like flavors.
The five flavors, $\Psi_u$, $\bar{\Psi}_d$ and $f$, $\bar{f}$, are charged under the standard model gauge symmetries
while the other $N_f = N+1$ flavors are SM singlets.
The matter content and charge assignment are summarized in Table~\ref{tab:model}.
The theory is in the conformal window, $3N/2 \leq N + 6 < 3N$.
To maintain the perturbative gauge coupling unification, the rank of the gauge group is constrained as
$N \lesssim 4$ with a vector-like pair of $\mathbf{{5}} + \mathbf{\bar{5}}$ messengers of gauge mediation
at an intermediate scale around 100\,TeV-1000\,TeV. 
In the case that the two pairs of the messengers exist, the perturbativity of the couplings holds up to $\sim 10^{15}$\, GeV with $N=4$, evaluated by two-loop renormalization group equations. The constraint may be relaxed if there is a large positive anomalous dimension, giving negative contributions to the beta-functions of the SM gauge couplings~\cite{syy}. 

We take the following superpotential in addition to the usual MSSM Yukawa couplings,
\begin{equation}
\begin{split}
W &= \lambda_u H_u \bar{\Psi}_d Q_{N_f} + \lambda_d H_d  {\Psi}_u \bar{Q}_{N_f} + \sum_I m_I Q_I \bar{Q}_I\\[1ex]
&\quad \hspace{-10pt} + \sum_{ij} \eta_{ij} X_{m} Q_i \bar Q_j  + \sum_A \Bigl(\eta_c^A Y_m \Phi_c^A \bar{\Phi}_c^A + \eta_l^A Y_m \Phi_l^A \bar{\Phi}_l^A    \\[1ex]
&\quad + M_c \Phi_c^A \bar{\Phi}_c^A + M_l \Phi_l^A \bar{\Phi}_l^A \Bigr)  \\[1ex]
&\quad+ m_{\Psi} {\Psi}_u \bar{\Psi}_d + m_f f \bar{f}  + M_{XY} X_m Y_m  + M_Y Y_m^2/2 \ ,
\label{Wmodel} 
\end{split}
\end{equation}
where $\lambda_u$, $\lambda_d$, $\eta_{ij}$ $(i,j = 1, \cdots , N_f-1, \ I=1, \cdots, N_f)$, ${\eta}_{c}^A$ and $\eta_l^A$
are dimensionless coupling constants and $m_\Psi$,  $m_f$, $M_{XY}$, $M_Y$, $m_I$, $M_c$ and $M_l$ are mass parameters. 
The fields $\Phi_c^A$, $\Phi_l^A$, $\bar \Phi_c^A$, and $\bar \Phi_l^A$ act as messenger superfields after the SUSY is broken, and $M_c$ and $M_l$ should be smaller than the confinement scale $\Lambda$, which will be described later. The index $A$ denotes the messenger number, and $A=1, \dots, N_5$.
We assume the following hierarchies among the mass parameters,
\begin{equation}
\begin{split}
m_{\Psi} \sim m_f \gg m_1> m_2 > \cdots > m_{N_f-1} \gg m_{N_f} .
\end{split}
\end{equation}
The last large hierarchy may be explained by imposing an approximate symmetry with a charge assignment, 
(charge of $m_{{Nf}}) \gg$ (charge of $m_i$). 
Also, couplings of the messenger fields to $Q_{N_f}$ and $\bar{Q}_{N_f}$ are forbidden by this approximate symmetry.

The theory is first in the conformal window and the hidden gauge coupling is in a fixed point at sufficiently high energies.
At the mass scale below $m_{\Psi} \sim m_f \sim M_{XY}$, ${\Psi}_u $, $\bar{\Psi}_d$, $f$, $\bar{f}$, $X_m$ and $Y_m$, are integrated out.
It is assumed that $M_Y \ll M_{XY}$.
Solving the equations of motion of these fields at classical level,
we obtain the following superpotential in the effective theory,
\begin{equation}
\begin{split}
W' &= - \frac{\lambda_u \lambda_d}{m_{\Psi}} Q_{N_f} \bar{Q}_{N_f} H_u H_d  \\
&\quad- \sum_{ij} \Bigl(\frac{\eta_{ij} \eta_c}{M_{XY}} Q_i \bar{Q}_j \Phi_c \bar{\Phi}_c
+ \frac{\eta_{ij} \eta_l}{M_{XY}} Q_i \bar{Q}_j \Phi_l \bar{\Phi}_l \Bigr)  \\
&\quad + \sum_{ijkl} \frac{\eta_{ij} \eta_{kl} M_Y}{2M_{XY}^2} Q_i \bar{Q}_j Q_k \bar{Q}_l  \\
&\quad+ \sum_I m_I Q_I \bar{Q}_I  + M_c \Phi_c \bar{\Phi}_c + M_l \Phi_l \bar{\Phi}_l, \label{Wprime}
\end{split}
\end{equation}
where the index $A$ is omitted here and hereafter. 
Decoupling of the five flavors, the effective theory is a SQCD with $(N + 1)$ flavors
where the gauge coupling gets strong and finally the theory (s-)confines.
The confinement scale $\Lambda$ is near the mass scale $m_{\Psi} \sim m_f $ when the gauge coupling is sufficiently strong before
the confinement.
Below the confinement scale, the hidden quarks, $Q_{I}$, $\bar{Q}_{J}$, form meson chiral superfields
and the first term of the above superpotential
leads to the NMSSM cubic coupling, $S H_u H_d$ ($S \sim Q_{N_f} \bar{Q}_{N_f}$), in a similar manner to the model of \cite{Chang:2004db}. Here, $M_Y \ll M_{XY}$ is favored to suppress masses of meson superfields.

The low-energy effective theory of a SQCD with $(N + 1)$ flavors has the following dynamically generated superpotential
as well as the superpotential Eq.~\eqref{Wprime},
\begin{equation}
\begin{split}
W_{\rm dyn} &= \frac{1}{\Lambda^{2N_f -3}} \left( \sum_{IJ} B_I M_{IJ} \bar{B}_J - \det M_{IJ} \right),
\end{split}
\end{equation}
where $M_{IJ} = Q_I \bar{Q}_J$ are meson chiral superfields and $B_I = \epsilon_{I I_1 \cdots I_{N}} Q_{I_1} \cdots Q_{I_{N}} / N !$,
$\bar{B}_I = \epsilon_{I I_1 \cdots I_{N}} \bar{Q}_{I_1} \cdots \bar{Q}_{I_{N}} / N !$ are (anti-)baryons.
We can rewrite these composite chiral superfields in terms of canonically normalized fields,
\begin{equation}
\begin{split}
S_{IJ} \sim \frac{M_{IJ}}{\Lambda} , \quad b_I \sim \frac{B_I}{\Lambda^{N_f - 2}}, \quad \bar{b}_I \sim \frac{\bar{B}_I}{\Lambda^{N_f - 2}},
\end{split}
\end{equation}
where the K\"ahler potential cannot be controlled and we have just put the dynamical scale $\Lambda$ so that
$S_{IJ}$, $b_I$ and $\bar{b}_I$ have mass dimension one correctly.
The total superpotential of the theory is then given by
\begin{equation}
\begin{split}
W_{\rm eff} &= \lambda_S S H_u H_d + M_c \Phi_c \bar{\Phi}_c + M_l \Phi_l \bar{\Phi}_l  \\[1ex]
&\quad+ \sum_{ij} ({\lambda}^c_{ij} S_{ij} \Phi_c \bar{\Phi}_c
+ {\lambda}^l_{ij} S_{ij} \Phi_l \bar{\Phi}_l) \\[1ex]
&\quad+  \sum_{IJ}\eta \, b_I S_{IJ} \bar{b}_J - \sum_I \eta' m_I \Lambda S_{II}- \eta'' \frac{\det S_{IJ}}{\Lambda^{N_f-3}} , \label{Weff}
\end{split}
\end{equation}
where $S \equiv S_{N_fN_f}$ and $\eta$, $\eta'$, $\eta''$ in front of the terms of the last line
are $\mathcal{O}(1)$ numerical factors. Mass terms, $S_{ij} S_{kl}$, will be discussed later.
The other coupling constants are estimated as
\begin{equation}
\begin{split}
&\lambda_S \sim  \frac{{\lambda_u \lambda_d}\Lambda}{m_{\Psi}} ,
\quad {\lambda}^c_{ij} \sim  \frac{{\eta_{ij} {\eta}_c}\Lambda}{M_{XY}},
\quad {\lambda}^l_{ij} \sim \frac{ \eta_{ij} {\eta}_l \Lambda}{M_{XY}}. \\[1ex]
\end{split}
\end{equation}
For $N_f = N+1 >3$, the last term $\sim \det S_{IJ}$ is irrelevant and not important around $S_{IJ} =0$.
We then ignore this term in the discussion of SUSY breaking below.

\subsection{Dynamical SUSY breaking}

We now show that dynamical SUSY breaking occurs at a meta-stable vacuum around the origin of the meson field space
in the present model.
There is a  local minimum at
\begin{equation}
\begin{split}
b = \bar{b} = \left( 
\begin{array}{cc}
\sqrt{m_1\Lambda} \\
0 \\
\vdots \\
0
\end{array}
\right),
\qquad S_{IJ} = 0,
\end{split}
\end{equation}
and the origin of the other fields in the theory.
We have not written the  $\mathcal{O}(1)$ numerical factors $\eta$, $\eta'$ explicitly.
The $F$-terms of the mesons $S_{IJ}$ at this minimum are given by
nonzero values,
\begin{equation}
\begin{split}
F_{S_{IJ}} = - \left( \frac{\partial W}{\partial S_{IJ}} \right)^\ast = m_{I} \delta_{IJ} \Lambda \neq 0 \qquad \text{for $I,J \neq 1$},
\label{F-term}
\end{split}
\end{equation}
which means SUSY is broken dynamically
\cite{Intriligator:2006dd}.
There is a supersymmetric vacuum far away from the origin of the meson field space,
but the lifetime of the meta-stable vacuum is sufficiently long if $m_1 \ll \Lambda$ is satisfied. 
With these $F$-terms, the gravitino mass is given by
\begin{eqnarray}
m_{3/2}^2 \simeq  \sum_{I\neq1} \frac{|m_I \Lambda|^2}{3 M_P^2},
\end{eqnarray}
imposing the condition for the vanishing cosmological constant. Here, $M_P (\simeq 2.43 \times 10^{18}$ GeV) is the reduced Planck mass.

Let us analyze the mass spectrum on the meta-stable vacuum.
Here, we ignore the small mass parameter $m_{N_f}$ just for simplicity.
We expand the baryons around the vacuum,
\begin{equation}
\begin{split}
b = &\left( 
\begin{array}{cc}
\sqrt{m_1\Lambda} + \delta \chi_1 \\
\delta \chi_2 \\
\vdots \\
\delta \chi_{N_f}
\end{array}
\right),
\quad \bar{b} = \left( 
\begin{array}{cc}
\sqrt{m_1\Lambda} + \delta \bar{\chi}_1 \\
\delta \bar{\chi}_2 \\
\vdots \\
\delta \bar{\chi}_{N_f}
\end{array}
\right).
\end{split}
\end{equation}
The supermultiplets of $S_{11}$, $S_{I1}$ and $S_{1I}$ ($I \neq 1$) form mass terms of order $\sqrt{m_1 \Lambda}$
with $(\delta \chi_1 + \delta \bar{\chi}_1)$,  $\delta \chi_I$,  $\delta \bar{\chi}_I$ respectively.
The supermultiplet of $(\delta \chi_1 - \delta \bar{\chi}_1)$ is the massless Nambu-Goldstone (NG) multiplet of the spontaneously broken
baryon number symmetry.
When we gauge this symmetry, we just obtain the massive Abelian gauge multiplet
by the super Higgs mechanism.
Alternatively, we can explicitly break the symmetry and make the NG mode massive.
The meson scalars $S_{IJ}$ ($I,J \neq 1$) are flat directions at classical level but
all of them except for the NMSSM singlet scalar $S$ get nonzero masses by the 1-loop Coleman-Weinberg potential.

The scalar of $S$ remains massless even at 1-loop level.
However, there is an important 2-loop correction for this scalar field,
which gives a runaway potential for the NMSSM singlet scalar.
Near the origin, the generated negative soft mass-squared is given by
\cite{Giveon:2008wp}
\begin{equation}
\begin{split}
V_{\rm 2-loop} \approx - 1.26 \, \eta^6 \frac{m_2 \Lambda}{(16\pi^2)^2}  \, S^\dagger S, \label{V2-loop}
\end{split}
\end{equation}
which makes it possible to explain the correct EWSB with the non-zero $\left<S\right>$, 
and then the Higgsino mass term is generated by the superpotential interaction $\lambda_S S H_u H_d$ as in the usual NMSSM. Here, $m_2 \gg m_3$ is assumed.
The potential \eqref{V2-loop} indicates that
the size of the soft mass depends on the SUSY breaking scale $\sqrt{m_2 \Lambda}$ 
while that tachyonic soft mass has to be smaller than the electroweak scale to preserve naturalness.
Then, if the coupling $\eta$ is $\mathcal{O}(1)$, $\sqrt{m_2 \Lambda}$ is not bigger than $\sim 100 \, \rm TeV$ in the present model. In this case, the cosmologically safe gravitino with a mass around 10\,eV is predicted.

We need to stabilize the runaway direction of the NMSSM scalar from the potential \eqref{V2-loop}.
In addition, some of the fermion components in $S_{IJ}$ are massless and need SUSY mass terms.
For these reasons, we consider higher dimensional operators,
\begin{equation}
\begin{split}
\Delta W = \frac{\lambda_{IJKL}}{M_0} Q_I \bar{Q}_J Q_K \bar{Q}_L = \tilde{m}_{IJKL} S_{IJ} S_{KL}, \label{higher}
\end{split}
\end{equation}
where $M_0$ is some ultra-violet (UV) mass scale and
\begin{equation}
\begin{split}
\tilde{m}_{IJKL} \sim \frac{{\lambda}_{IJKL}\Lambda^2}{M_0}.
\end{split}
\end{equation}
Here, among the higher dimensional operators, $W \ni Q_i\bar Q_j Q_k \bar Q_l$ arises from Eq.\,(\ref{Wmodel}). 
The higher dimensional operators in Eq.\,(\ref{higher}) can give nonzero masses for all the meson scalars and fermions except for the massless goldstino of SUSY breaking. The mass terms 
\begin{eqnarray}
\tilde m_{N_f N_f p p}, \ \tilde m_{pp N_f N_f} \,\, (p=,2 \dots, N_f-1) 
\end{eqnarray}
should be suppressed, since otherwise the EWSB scale is destabilized by $\Delta V \sim  F_{pp} (\tilde m_{pp N_f N_f} + \tilde m_{N_f N_f pp})S + {\rm h.c.}$ We can suppress these dangerous mass terms by 
imposing a symmetry which forbids $(Q_{N_f} \bar Q_{N_f})(Q_{p} \bar Q_{q})$, where $q=2, \dots, N_f-1$.

In addition, the linear terms $\Delta V \sim F_{pp} (\tilde m_{pp q r} + \tilde m_{qr pp } )S_{qr}  + {\rm h.c.}$ shift the vacuum expectation values of the pseudo-moduli mesons from the origin,
$\Delta S_{qr} \sim 16 \pi^2 \Lambda^2 / M_0 \sim 16 \pi^2 \tilde{m} $ as discussed in \cite{Murayama:2006yf,Kitano:2006xg}.
This must be smaller than $4\pi \sqrt{m_2 \Lambda}$ so that the analysis of SUSY breaking above is valid
and also smaller than the messenger masses $M_c \sim M_l$ for the messengers not to be tachyonic (See below).
Consequently, $\tilde m_{pp qr}$ and $\tilde m_{qr pp}$ can not be significantly larger than 0.1-1 TeV.

\renewcommand{\arraystretch}{1.3}
\begin{table*}[!t]
\begin{center}
\begin{tabular}{|c|cccccccccccc|}
\hline
 &  $Q_{N_f}$ & $\bar{Q}_{N_f}$ & $Q_{p}$ & $\bar{Q}_{q}$ & $X_1$ & $X_2$ & $X_3$ &$X_m$ & $Y_m$ & $(H_u \bar \Psi_d)$ & $(H_d \Psi_u)$ & $(\Phi_{l,c} \bar \Phi_{l,c})$ \\
 \hline
  $U(1)$  &  $-4/5$ & $-1/5$ & $-1/5$ & $0$ & $1$ & $4/5$ & $2/5$ & $1/5$  &$-1/15$ & 4/5 & 1/5 &$1/15$ \\
\hline
\hline
 &  $M_{1}$ & $M_{2}$ &  $M_{3}$ &  $M_{XY}$ & $m_{N_f}$ & $m_{p}$ & $M_Y$ & $M_{l,c}$ & &   &&  \\
 \hline
  $U(1)$  &  $-2$ & $-8/5$ & $-4/5$ & $-2/15$ &  1 & 1/5 & 2/15 &$-1/15$ &&&&   \\
\hline
\end{tabular}
\end{center}
\caption{The example of the $U(1)$ charge assignment. 
The mass parameters $m_{N_f}$, $m_{p}$, $M_{XY}$, $M_Y$ and $M_{a} (a=1,2,3)$ are considered to be spurions of $U(1)$ symmetry breaking.
(Also, $m_{\Psi}$ ($m_{f}$) has a non-vanishing charge.)
}
\label{tab:PQ}
\end{table*}
\renewcommand{\arraystretch}{1}

The higher dimensional operators \eqref{higher} to satisfy the constraints can be UV completed
in the following renormalizable superpotential for instance,
\begin{eqnarray}
W &=& X_1 Q_{N_f} \bar Q_{N_f} + \sum_p ( X_2 Q_{N_f} \bar Q_{p} +  X_3 Q_{p} \bar Q_{N_f} ) \nonumber \\
&+& \sum_{p,q} X_m (Q_p \bar Q_q) + M_{XY} X_m Y_m  + \frac{1}{2} M_Y Y_m^2\nonumber \\
&+& \frac{1}{2} (M_1 X_1^2 + M_2  X_2^2 + M_3 X_3^2 ) \,,
\end{eqnarray}
where $p,q = 2, \dots, N_f-1$, and $X_a$ ($a=1,2,3$) are chiral superfields which are singlet under the standard model and $SU(N)_H$ gauge symmetries. The above superpotential is explained by an approximate $U(1)$ symmetry, with 
a charge assignment summarized in Table~\ref{tab:PQ}.
A Yukawa coupling is implicitly multiplied by each of cubic terms.
Linear terms of $X_1$, $X_m$ and $Y_m$ have been removed by shifts of the fields.
The charge of $(H_u H_d)$ is chosen to be non-zero such that the bare mass term, 
so-called $\mu$-term, is prohibited. 
The MSSM matter fields, which are not shown here,  are also charged under this approximate symmetry.

When new fields $X_i$ are integrated out, we obtain the higher dimensional operators \eqref{higher}
with forbidden $\lambda_{pq N_fN_f}$ and $\lambda_{N_fN_f pq}$,  correctly. (Therefore, $\tilde m_{N_f N_f pp}$ 
and $\tilde m_{pp N_f N_f}$  vanish.) Other mass terms which mix the NMSSM singlet $S$ to other mesons 
e.g. $m_{N_f N_f p N_f}$ are prohibited.
Since the symmetry forbids the messengers-$Q_{N_f}$ ($\bar{Q}_{N_f}$) couplings such as $\Phi_l \bar{\Psi}_d Q_{N_f}$
in the superpotential Eq.\,\eqref{Wmodel}, couplings of $S$ to the messenger pairs, $S \Phi_c \bar{\Phi}_c$ and $S \Phi_l \bar{\Phi}_l$, vanish: the EWSB scale is not destabilized.

\subsection{Gauge mediation}

We now consider gauge mediation of SUSY breaking to ordinary superparticles.
Integrating out the messenger fields, the gaugino mass and scalar masses are generated as
\begin{equation}
\begin{split}
m_{\rm soft} \sim \frac{g^2}{16 \pi^2} \frac{\bar{m} \Lambda}{M}, \label{msoft}
\end{split}
\end{equation}
where $g$ denotes the standard model gauge coupling and we have defined $\bar{m} \equiv \sum_{i \neq 1} \lambda_{ii} m_i$. Here, $M \sim M_c \sim M_l$ is the messenger mass scale.
To obtain the soft masses at the electroweak scale, we set
\begin{equation}
\begin{split}
\Lambda_{\rm mess} \equiv \frac{\bar{m} \Lambda}{M} \sim 100 \, \rm TeV .
\end{split}
\end{equation}
The condition that the messenger fields are not tachyonic gives ${\bar{m} \Lambda} < M^2$.
The messengers also give the 1-loop Coleman-Weinberg potential to the meson scalars $S_{ij}$,
which also leads to the shifts of the pseudo-moduli
\cite{Murayama:2006yf}.
Since there is a vacuum with lower energy where the messenger scalars are condensed,
a tachyonic direction appears around $S_{ij} \sim M$
(The existence of this vacuum is essential for non-vanishing leading order gaugino masses.
See e.g. \cite{Kitano:2006xg,Komargodski:2009jf,Nakai:2010th}).
Then, the stabilized point of the pseudo-moduli has to be smaller than the messenger mass scale $M$,
which gives a constraint, $M > \lambda^2 \sqrt{m \Lambda}$.
Note that the transition between the SUSY breaking local minimum to the minimum with $S_{ij} \sim M$  provides a similar but more stringent constraint, $M \gtrsim 3 \lambda \sqrt{m \Lambda}$~\cite{hisano_vac},\footnote{
If the messenger superfields are not thermalized,  the constraint becomes weaker as
$M \gtrsim 1.5 \lambda \sqrt{m \Lambda}$, provided that the SUSY breaking local minimum is selected in the early universe.
} which will be discussed in the next section.

\section{Phenomenology}\label{sec:discuss}

We now turn to discuss the phenomenology of the model. As mentioned in the previous section, gaugino, squark and slepton masses are the same as those in usual gauge mediation with a messenger mass $M$ and $B$-term of $\bar{m} \Lambda= M \Lambda_{\rm mess}$. On the other hand, values of soft breaking parameters in the extended Higgs sector are different from those of usual gauge mediation scenarios.

The low-energy effective superpotential of the extended Higgs sector can be written as 
\begin{equation}
\begin{split}
W &\supset \xi_F S + \frac{1}{2} \mu^\prime S^2 + \lambda_S S H_u H_d, \label{eq:nmssm1}
\end{split}
\end{equation}
where $\xi_F = - m_{N_f} \Lambda$ in Eq.~\eqref{Weff} and $\mu^\prime = 2 \tilde{m}_{N_f N_f N_f N_f}$ in Eq.~\eqref{higher}, thus, we treat these and $\lambda_S$ as free parameters here. 
Note that the large $\lambda_S$ of $\sim 1$ is quite natural as shown in Appendix B.

The coupling of $S^3$ term will be suppressed because it is provided from a higher dimensional operator, $(Q_{N_f} \bar Q_{f})^3$, and hence, it has been neglected. The corresponding SUSY breaking terms are defined as 
\begin{equation}
\begin{split}
V &\supset \left( \xi_S S + \frac{1}{2} m_S^{\prime 2} S^2 + A_\lambda \lambda_S S H_u H_d + {\rm h. c.} \right) \\ 
&\quad+ m_S^2 |S|^2 + m_{H_u}^2 |H_u|^2 + m_{H_d}^2 |H_d|^2, \label{eq:nmssm2}
\end{split}
\end{equation}
where the $m_S^2$ receives the negative two-loop contribution $\sim \eta^6 m_2 \Lambda / (16 \pi^2)^2 \sim \eta^6 (m_{3/2} M_P)/(16 \pi^2)^2$ written in Eq.~\eqref{V2-loop}. 
It turns out that this negative $m_S^2$ is important for the successful EWSB. 
%
%
The other soft mass parameters are
\begin{equation}
\begin{split}
A_\lambda &\approx 0 ~{\rm GeV}, \quad 
\xi_S \approx 0 ~{\rm GeV}, \quad 
m_S^\prime \approx 0 ~{\rm GeV}, 
\end{split}
\end{equation}
at the messenger scale because the singlet $S$ does not couple messenger fields directly. 
Hereafter we assume all parameters are real for simplicity.

\subsection{EWSB and Higgs mass}

To show the viable parameter space and typical mass spectra, we discuss the feature of the extended Higgs sector. We define neutral scalar components of Higgs as $H_u^0 =v_u + ( h_u + i a_u )/\sqrt{2}$, $H_d^0 = v_d + ( h_d + i a_d )/\sqrt{2}$ and $S = v_S + ( s_R + i s_I )/\sqrt{2}$, respectively. These $v_u, v_d, v_S$ denote vacuum expectation values and $\tan\beta \equiv v_u/v_d$ and $v \equiv \sqrt{v_u^2 + v_d^2} \simeq 174.1$ GeV. The vacuum conditions can be obtained by 
\begin{equation}
\begin{split}
\frac{m_Z^2}{2}
&\approx - \mu_{\rm eff}^2 + \frac{m_{H_d}^2 - m_{H_u}^2 \tan^2\beta}{\tan^2\beta - 1}, 
\\
\sin2\beta 
&\approx \frac{2(\mu_{\rm eff} \mu^\prime + \lambda_S \xi_F)}
           {m_{H_u}^2 + m_{H_d}^2 + 2\mu_{\rm eff}^2 + \lambda_S^2 v^2}, 
\\ 
\mu_{\rm eff}
&\approx -\frac{\lambda_S \mu^\prime}{2} 
          \frac{2 \xi_F - \lambda_S v^2 \sin2\beta}{m_S^2+\mu^{\prime 2}+\lambda_S^2 v^2},
\label{eq:vacuumcondition}
\end{split}
\end{equation}
at tree level. Here $m_Z^2 = (g_Y^2 + g_2^2)v^2/2$ and $\mu_{\rm eff} \equiv \lambda_S v_S$. 
Note that without the negative $m_S^2$, $\mu_{\rm eff}$ is predicted to be around $-\lambda_S \xi_F/\mu'$. 
As a result, the predicted value of $\tan\beta$ is huge: the successful EWSB does not occur, 
unless the Lagrangian with Eq.\,(\ref{eq:nmssm1}) and (\ref{eq:nmssm2}) becomes a MSSM limit by $\lambda_S \to 0$ and $(\lambda_S \xi_F/\mu')=$\,fixed.

\begin{figure}[!t]
\begin{center}
\includegraphics[scale=0.85]{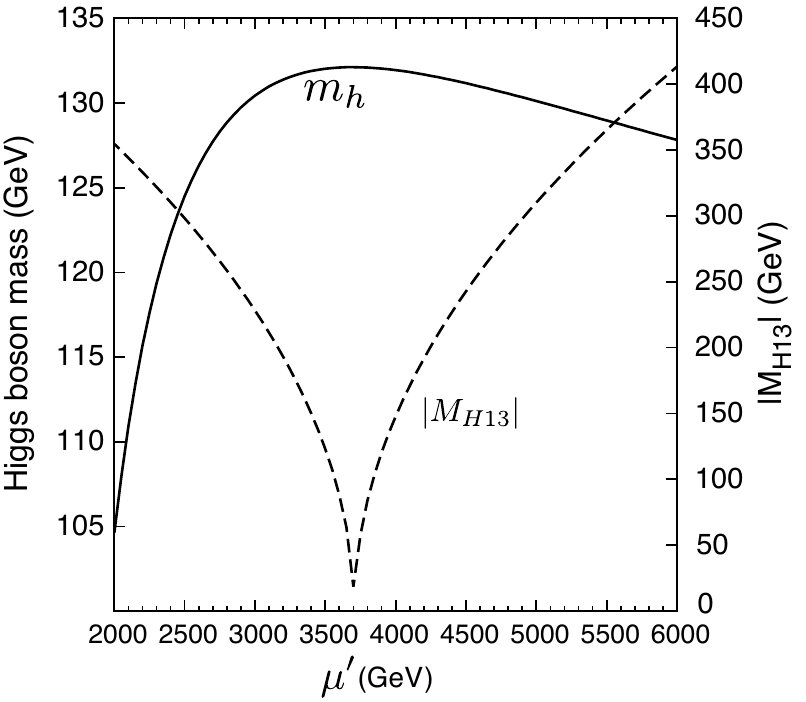}
\caption{
The lightest Higgs boson mass $m_h$ and an off-diagonal component of the neutral CP-even Higgs mass matrix $M_{H13}$. We take ($\lambda_S, \tan\beta, \mu_{\rm eff}/|\mu_{\rm eff}|) = (1.0, 4.0, 1)$ and $(M, \Lambda_{\rm mess}, N_5)=(300\,{\rm TeV}, 180\,{\rm TeV}, 1)$. Here, $\alpha_s(M_Z)=0.1185$ and $m_t({\rm pole})=173.34$ GeV.
}
\label{fig:mixing}
\end{center}
\end{figure}

\begin{figure}[!t]
\begin{center}
\includegraphics[scale=1.0]{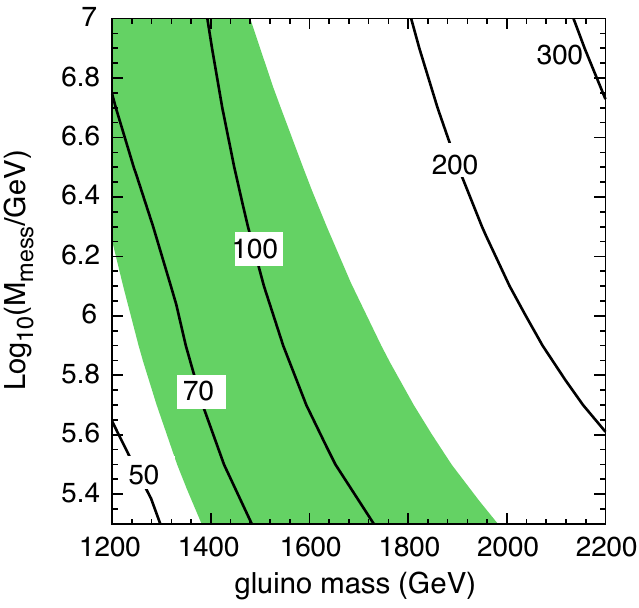}
\caption{
The fine-tuning measure $\Delta$ (black solid). We take $\lambda_S=1.0$, $\tan\beta=4$, $\mu_{\rm eff}>0$, $\mu'=6$\,TeV and $N_5=2$, while the other parameters are the same as in Fig.~\ref{fig:mixing}. In the green (shaded) region, the lightest Higgs boson mass is in a range of 122-128 GeV. 
}
\label{fig:mh_delta}
\end{center}
\end{figure}

Using the following base 
\begin{eqnarray}
\left(
\begin{array}{c}
      h_1^\prime \\
      h_2^\prime \\ 
      h_3^\prime \\
\end{array}
       \right) 
&=& \left(
\begin{array}{ccc}
      \cos\beta & -\sin\beta & 0\\
      \sin\beta & \cos\beta  & 0\\ 
      0         & 0          & 1\\ 
\end{array}
       \right)
\left(
\begin{array}{c}
      h_d \\
      h_u \\ 
      s_R \\
\end{array}
       \right),
  \label{eq:rotationHiggsfield}
\end{eqnarray}
the CP-even Higgs mass squared matrix can be written as 
\begin{eqnarray}
M_{\rm H}^2
&=& \left(
\begin{array}{ccc}
      M_{\rm H11}^2 & M_{\rm H12}^2 & M_{\rm H13}^2\\
                        & M_{\rm H22}^2 & M_{\rm H23}^2\\ 
                        &                   & M_{\rm H33}^2\\ 
\end{array}
       \right),
  \label{eq:Higgsmassmatrix1}
\end{eqnarray}
where 
\begin{equation}
\begin{split}
M_{\rm H11}^2 &\approx 
m_Z^2 (\cos2\beta)^2 + \lambda_S^2 v^2 (\sin2\beta)^2, \\ 
M_{\rm H22}^2 &\approx 2 (\mu_{\rm eff} \mu^\prime + \lambda_S \xi_F) / \sin2\beta 
\\ & \quad +\left(m_Z^2 - \lambda_S^2 v^2 \right) (\sin2\beta)^2, \\ 
M_{\rm H33}^2 &\approx \mu^\prime(- \lambda_S \xi_F + \lambda_S^2 v^2 \sin\beta/2)/\mu_{\rm eff}, \\ 
M_{\rm H12}^2 &\approx (- m_Z^2 + \lambda_S^2 v^2)\sin4\beta/2, \\ 
M_{\rm H13}^2 &\approx \lambda_S v ( 2 \mu_{\rm eff} - \mu^\prime \sin2\beta), \\ 
M_{\rm H23}^2 &\approx -\lambda_S v \mu^\prime \cos2\beta, 
\label{eq:Higgsmassmatrix2}
\end{split}
\end{equation}
at tree level. 
As we can see, there are positive and negative contributions to the lightest Higgs boson mass: A positive contribution is the additional $F$-term in $M_{\rm H11}^2$, $\lambda_S^2 v^2 (\sin2\beta)^2$ and a negative contribution comes from $M_{\rm H13}^2$ after diagonalizing the mass squared matrix. Therefore, small $\tan\beta$ and $\mu_{\rm eff} \sim \mu^\prime \sin2\beta/2$ are favored to push the lightest Higgs boson mass up.

Actually, we can see that the lightest Higgs boson mass is maximized around the minimum of $|M_{\rm H13}|$ in Fig.~\ref{fig:mixing}. In our numerical calculations, the Higgs boson mass and SUSY mass spectra are calculated using {\tt NMGMSB}~\cite{Ellwanger:2008py}, which is contained in {\tt NMSSMTools 4.8.2}. The NMSSM parameters are taken as $(\lambda_S, \tan\beta, \mu_{\rm eff}/|\mu_{\rm eff}|)=(1.0, 4.0, 1)$, while parameters in the messenger sector are $(\Lambda_{\rm mess}, M, N_5)$ = (180\,TeV, 300\,TeV, 1), where $N_5$ is a number of the messenger pairs.
%
%
%
%
%
It should be noted that considering above discussion on the Higgs boson mass and vacuum conditions in Eq.~\eqref{eq:vacuumcondition},　
the values of $\mu_{\rm eff}$, $\mu^\prime$, $\sqrt{\xi_F}$ and $m_S$ would be the same order in the viable parameter space.

\vspace{12pt}
We also estimate the fine-tuning of the EWSB scale using the following fine-tuning measure~\cite{ft_measure}:\,\footnote{
The definition of $\Delta$ here differs from the original one by a factor 2.
With the definition of Eq.\,(\ref{eq:ftm}), $\partial \ln v /\partial \ln |\mu| \simeq 2 \mu^2/m_Z^2$ in the MSSM. 
}
\begin{eqnarray}
\hspace{-5pt}\Delta = {\rm max} \Bigl\{ \left|\frac{\partial \ln v}{\partial \ln |a|} \right|
 \Bigr\}, \left(a \in　
\begin{array}{c}
{\rm fundamental\,\, mass} \\
{\rm parameters} 
\end{array}\right) \label{eq:ftm}
\end{eqnarray}
where $a=\xi_F, \mu'$, $\Lambda_{\rm mess}$, $|m_S^2|$ in our case.
($d \ln |\xi_F|$, $d \ln |\Lambda_{\rm mess}|$ and $d \ln |m_S^2|$ correspond to $d \ln |m_{N_f}|$, $d \ln |\bar m|$ and $d \ln |m_2|$, respectively.)

\vspace{5pt}
In Fig.~\ref{fig:mh_delta}, the fine-tuning measure $\Delta$ is shown on the gluino mass (pole mass)-$M$ plane. Within the green (shaded) region, the lightest Higgs boson mass is in a range of 122-128 GeV.  It is noticed that, thanks to the additional $F$-term contribution, the observed Higgs boson mass around 125 GeV is easily explained even with the 1.2 TeV gluino mass in the gauge mediation scenario, as pointed out in Ref.~\cite{nmssm_gmsb_higgs}. 
The larger gluino mass can of course also be consistent with the observed Higgs boson mass by changing the $\lambda_S$ value to slightly smaller. As a result, the fine-tuning of the EWSB scale is drastically improved, compared to the MSSM in gauge mediation; $\Delta \lesssim 100$ for $M \simeq 200$ TeV.

\begin{table*}[]
\begin{tabular}{|c||c|c|c|c|}
\hline
Parameters & Point {\bf I} & Point {\bf II} \\
\hline
$\Lambda_{\rm mess}$ (TeV) & 185  & 130  \\
$M$ (TeV)& 1000 & 200  \\
$N_5$ & 1  & 3   \\
\hline
$\tan\beta$ & 4.5 & 5.0  \\
$\lambda_S$ & 0.9 & 0.92  \\
$\mu_{\rm eff}$ (GeV) & 783 & 893  \\
$\mu^\prime$ (GeV) & 6000 & 4000  \\
$\xi_F$ (GeV)$^2$ & -4.5$\times10^6$ & -3.3$\times10^6$  \\
$m_S^2$ (GeV)$^2$ & -1.95$\times10^6$ & -1.41$\times10^6$  \\
\hline
$\Delta$ & 165 & 235 \\
\hline
Particles & Mass (GeV) & Mass (GeV) \\
\hline
$\tilde{g}$ & 1430 & 2890  \\
$\tilde{q}$ & 1840-1940 & 2570-2680  \\
$\tilde{t}_{2,1}$ & 1860, 1660 & 2610, 2410  \\
$\tilde{\chi}_{2,1}^\pm$ & 807, 486 & 1130, 886  \\
$\tilde{\chi}_5^0$ & 5160 & 3610  \\
$\tilde{\chi}_4^0$ & 807 & 1130  \\
$\tilde{\chi}_3^0$ & 796 & 910  \\
$\tilde{\chi}_2^0$ & 486 & 886  \\
$\tilde{\chi}_1^0$ & 253 & 580  \\
$\tilde{e}_{L, R}(\tilde{\mu}_{L, R})$ & 652, 326.6 & 810, 392.9  \\
$\tilde{\tau}_{2,1}$ & 652, 325.9 & 810, 392.3 \\
$H^\pm$ & 851 & 1137  \\
$A_{2, 1}$ &  4940, 847 & 3380, 1133  \\
$h_{1}$ & 124.7 &  125.1 \\
\hline
\end{tabular}
\caption{\small Mass spectrum in sample points. Values of $\lambda_S$ and $\tan\beta$ show the values at the SUSY scale and values of $\xi_F$, $\mu^\prime$ and $M_S^2$ show the values at the messenger scale in this table.
Here, $A_1$ ($A_2$) denotes the lighter (heavier) CP-odd Higgs.
}
\label{tab:sample point}
\end{table*}


\subsection{Cosmology}

Before closing this section, let us discuss cosmological aspects and implications to the collider signals. In our model, 
the gravitino mass is estimated as 
\begin{eqnarray}
m_{3/2} &\simeq& \frac{\bar m \Lambda}{\lambda} \frac{1}{\sqrt{3} M_P} \nonumber \\
&\approx& 7\, {\rm eV} \cdot 
\left(\frac{\Lambda_{\rm mess}}{100\, {\rm TeV}}\right)
\left(\frac{M}{300\, {\rm TeV}} \right)\frac{1}{\lambda}, 
\end{eqnarray}
where $\lambda$ is a typical value of $\lambda_{ii}$, provided $\lambda_{ii} \sim \lambda_{jj} (i \neq j)$.

To satisfy the warm dark matter constraint, 
the gravitino should be lighter than 16\,eV~\cite{gravitino_wdm}, and in this range there is no constraint on the reheating temperature.
Then, the messenger scale is bounded from above as
\begin{eqnarray}
M < 6.7 \times 10^5 \,{\rm GeV} \cdot \lambda \left(\frac{100 {\rm TeV}}{\Lambda_{\rm mess}}\right) 
\label{eq:stability1},
\end{eqnarray}
with
\begin{eqnarray}
\lambda \gtrsim 0.15 \cdot \left(\frac{\Lambda_{\rm mess}}{100\,{\rm TeV}}\right)^2, \label{eq:stability1}
\end{eqnarray}
which is required not to conflict with $M^2 > \bar m \Lambda$.
On the other hand, to avoid the unstable SUSY breaking minimum with a life-time shorter than the age of the universe, 
the messenger scale should be~\cite{hisano_vac}
\begin{eqnarray}
M \gtrsim  9 \times 10^5 {\rm GeV} \cdot  \, \lambda^2 \left( \frac{\Lambda_{\rm mess}}{100\,{\rm TeV}} \right).\label{eq:stability2}
\end{eqnarray}
Combining the above three conditions, we get the upper-bound on $\Lambda_{\rm mess}$ as
\begin{eqnarray}
\Lambda_{\rm mess} \lesssim 150\, {\rm TeV}.
\end{eqnarray}
For $N_5=1$, it is difficult to satisfy this upper-bound taking into account null results of latest LHC SUSY searches.\footnote{
It may be possible to avoid this constraint if the reheating temperature 
is sufficiently low or the gravitino abundance is diluted by the late-time entropy production, allowing the larger gravitino mass than 16\,eV.
} By demanding that the gluino mass, $m_{\tilde g}$, be larger than 1.4 (1.6) TeV, 
the lower-bound on $\Lambda_{\rm mess}$ is 
\begin{eqnarray}
\Lambda_{\rm mess} \gtrsim 180\, (205) \,{\rm TeV}/N_5,
\end{eqnarray}
for $\Lambda_{\rm mess}/M = 1/2$. 
Therefore, $N_5 > 1$ is required. With the very light gravitino, the next-to-lightest SUSY particle (NLSP), which is likely to be the stau in our model, decays promptly~\cite{Asano:2011ri}. In our case, right-handed sleptons are almost degenerated in mass, and the strong constraint comes from the SUSY searches in final states with multi-jets, multi-leptons ($\gtrsim 3$) and missing transverse momentum~\cite{Chatrchyan:2014aea}. The chargino needs to be heavier than 850-900 GeV, resulting in the lower-bound on the gluino mass as $m_g > 2.4$-\,$2.7$ TeV with a GUT relation among gaugino masses. Here, $m_{\tilde g}$ is the gluino mass.
%

In the meantime, if the gravitino mass is in a range
\begin{eqnarray}
m_{3/2} \gtrsim \mathcal{O}(10)\,{\rm keV},
\end{eqnarray}
the gravitino can be a cold dark matter~\cite{gravitino_cdm} with an appropriate reheating temperature or the late-time entropy production~\cite{gmsb_entropy}. The lower bound on the gravitino mass in turn leads to the lower bound on the messenger scale:
\begin{eqnarray}
M> 4.2 \times 10^8 {\rm GeV} \cdot \lambda \left( \frac{m_{3/2}}{10\,{\rm keV}} \right) \left( \frac{100\,{\rm TeV}}{\Lambda_{\rm mess}} \right),
\end{eqnarray}
or equivalently 
\begin{eqnarray}
(\bar m \Lambda/\lambda)^{1/2} \gtrsim 2.1\times 10^8\, {\rm GeV} \left(\frac{m_{3/2}}{10\,{\rm keV}} \right)^{1/2}.
\end{eqnarray}
In this case, the stability bound of the SUSY breaking minimum, Eq.\,(\ref{eq:stability2}), is no longer important. 
Note that the messenger scale can be still low as $\sim 10^6$ GeV for $\lambda \sim 0.01$, and $\eta\sim 0.2$ (see Eq.~(\ref{Weff})). 

Since the typical decay length of the NLSP is $\mathcal{O}(10)$\,m, the stau NLSP is strongly constrained as $m_{\tilde \tau_1} \gtrsim $450-500\, GeV~\cite{stable_stau, ATLAS:2014fka}, where $m_{\tilde \tau_1}$ is the (lighter) stau mass. Therefore, the bino NLSP predicted with $N_5=1$ may be favored if the gravitino is the cold dark matter. In this case, the LHC signatures as well as limits of SUSY particle masses are similar to those in gravity mediation, and the strongest constraint comes from the SUSY searches with multi-jets and missing transverse momentum. So far, it is expected that $m_{\tilde g} \gtrsim $1.4-1.6\,TeV is required~\cite{atlas_multijets}, depending on the squark mass. 

\vspace{12pt}
Finally, we show the typical mass spectra and fine-tuning measure $\Delta$ in Table~\ref{tab:sample point}. 
At the point {\bf I} with $N_5=1$, the bino-like neutralino is the NLSP and $m_{3/2} \gtrsim \mathcal{O}(10)$\,keV which enables us to explain the observed dark matter by the gravitino. As discussed above, the gravitino mass with $m_{3/2}<16$ eV is difficult to be achieved in this case. 
On the other hand, at the points ${\bf II}$, the stau is the NLSP; therefore, $m_{3/2}<16$ eV is required to avoid the strong constraint on the stable stau without increasing the SUSY mass scale. In such a very light gravitino region, there is no constraint on the reheating temperature and the observed dark matter relic would be explained by another particle (e.g. QCD axion). 

\vspace{0pt}

\section{Conclusion and Discussion}\label{sec:discuss}

We have proposed a new scheme for the NMSSM in gauge mediation, 
where in general the successful EWSB does not occur due to the absence of the soft SUSY breaking mass parameter for $S$.  In our framework, $S$ is a composite meson in the hidden QCD, which is responsible for the dynamical SUSY breaking. The required soft SUSY breaking mass for $S$ naturally arises after SUSY is broken dynamically, and the electroweak symmetry is successfully broken. 

Although the UV Lagrangian Eq.~(\ref{Wmodel}) contains two singlets, it is possible to construct a model without singlets as shown in Appendix A, where the required particle content becomes less. In this case, there are additional contributions to the soft masses for the singlet and Higgs doublets, and all trilinear couplings. The stop, sbottom and stau masses are also modified. Therefore, the quite different phenomenology is expected to appear.


As a concrete model of the dynamical SUSY breaking, we have utilized the ISS model. 
However, it is possible to consider other dynamical SUSY breaking models. 
For instance, if the IYIT model \cite{Izawa:1996pk,Intriligator:1996pu} is adopted, a model similar to the Dirac NMSSM~\cite{dirac_nmssm} appears as a low-energy effective theory~\cite{in_pre}, which also increases the Higgs boson mass in a similar but different way. It needs to be checked whether the correct EWSB is explained. 

In our model, the gravitino can be either very light as $\sim 10$\,eV or $\mathcal{O}(10)$\,keV. 
In the former case, there is no upper bound for the reheating temperature, but the gravitino can not explain the observed dark matter abundance. One needs another candidate for a dark matter. 
For instance, the stable baryon in the hidden QCD may explain the observed abundance of the dark matter.  
In the latter case, the gravitino is cold enough and a dark matter candidate.  Although the gravitino tends to be over-produced in general, it is possible to fit for the standard cosmology if the late-time entropy production exists~\cite{gmsb_entropy}.

\vspace{0pt}

\section{Acknowledgments}\label{sec:ackno}

This work is supported by the German Research Foundation through TRR33 ``The Dark Universe" (MA). YN is supported by a JSPS Fellowship for Research Abroad. The research leading to these results has received funding from the European Research Council under the European Unions Seventh Framework Programme (FP/2007-2013) / ERC Grant Agreement n. 279972 ``NPFlavour'' (NY).

\appendix

\section{UV model without singlets}

An UV model without singlets is shown. The particle contents are less than Eq.~(\ref{Wmodel}). The superpotential is
\begin{equation}
\begin{split}
W &= \lambda_u H_u \bar{\Psi}_d Q_{N_f} + \lambda_d H_d  {\Psi}_u \bar{Q}_{N_f} \\[1ex]
&\quad \hspace{-10pt} + \sum_A \Bigl( \eta_i^c \Phi_c^A \bar{f} Q_i + \bar{\eta}_i^c \bar{\Phi}_c^A {f} \bar{Q}_i  
+ \eta_i^l \Phi_l^A \bar{\Psi}_d Q_i + \bar{\eta}_i^l \bar{\Phi}_l^A {\Psi_u} \bar{Q}_i \\[1ex]
&\quad + M_c \Phi_c^A \bar{\Phi}_c^A + M_l \Phi_l^A \bar{\Phi}_l^A
\Bigr) \\[1ex]
&\quad+ m_{\Psi} {\Psi}_u \bar{\Psi}_d + m_f f \bar{f}  + m_I Q_I \bar{Q}_I.
\end{split}
\end{equation}
After integrating out $f$, $\bar f$, $\Psi_u$ and $\bar \Psi_d$, the number of the flavor becomes $N_f=N+1$ and the theory confines. Then, the low-energy effective Lagrangian is
\begin{equation}
\begin{split}
W_{\rm eff} &= \lambda_S S H_u H_d + M_c \Phi_c \bar{\Phi}_c + M_l \Phi_l \bar{\Phi}_l  \\[1ex]
&\quad+ {\lambda}^c_{ij} S_{ij} \Phi_c \bar{\Phi}_c
+ {\lambda}^l_{ij} S_{ij} \Phi_l \bar{\Phi}_l \\[1ex]
&\quad+ \tilde{\lambda}^u_{i} S_{N_f i} H_u \bar{\Phi}_l
+ \tilde{\lambda}^d_{i} S_{iN_f} H_d \Phi_l \\
&\quad+  \eta \, b_I S_{IJ} \bar{b}_J - \eta' m_I \Lambda S_{II}- \eta'' \frac{\det S_{IJ}}{\Lambda^{N_f-3}}. 
\end{split}
\end{equation}
In this model, there are messenger-Higgs couplings, $S_{N_f i} H_u \bar{\Phi}_l$ and $S_{iN_f} H_d \Phi_l$. These couplings generate two-loop negative contributions to $m_S^2$ other than the contribution shown in Eq.\,(\ref{V2-loop}). 
The soft masses of $H_u$ and $H_d$ as well as those of the stop, sbottom and stau are also modified by two-loop effects. The trilinear couplings $A_\lambda$, $A_t$, $A_b$ and $A_{\tau}$ are generated at the one-loop level, where $A_t$, $A_b$ and $A_{\tau}$ are those of the stop, sbottom, and stau, respectively. Size of one-loop corrections to $m_{H_u}^2$ and $m_{H_d}^2$ are parametrically similar to those of two-loop contributions. These one-loop corrections to  $m_{H_{u,d}}^2$ vanish for $\Lambda_{\rm mess}/M \to 0$.

\section{Couplings at the fixed point}

In this Appendix, we estimate the couplings relevant to $\lambda_S$.
Provided that $\lambda_u H_u \bar{\Psi}_d Q_{N_f}$, $\lambda_d H_d  {\Psi}_u \bar{Q}_{N_f}$ and $\eta_{ij} X_{m} Q_i \bar Q_j$ in Eq.\,(\ref{Wmodel}) are at the fixed point, using the $a$-maximization technique~\cite{Intriligator:2003jj},
the anomalous dimensions of the fields are  
\begin{eqnarray}
&& \gamma_{H_u} =\gamma_{H_d} = 0.148, \ \ 
\gamma_{\Psi_u} =\gamma_{{\bar \Psi}_d} = -0.088, \nonumber \\\ \
&& \gamma_{f} =\gamma_{\bar f} = -0.114, \ \ 
\gamma_{Q_{N_f}} =\gamma_{{\bar Q}_{N_f}} = -0.060, \ \ \nonumber \\ 
&& \gamma_{Q_i} =\gamma_{{\bar Q}_i} = -0.106, \ \ 
\gamma_{X_m} = 0.211,
\end{eqnarray}
for $N=4$.
On the other hand, the one-loop calculations give
\begin{eqnarray}
&&\gamma_{H_u} = \frac{1}{16\pi^2} ( N \lambda_u^2),  \ \ 
\gamma_{H_d} = \frac{1}{16\pi^2} ( N \lambda_d^2), \ \ \nonumber \\
&&\gamma_{X_m} = \frac{1}{16\pi^2} ( N_f-1) N  \eta_Q^2 \ ,  
\end{eqnarray}
where $\eta_{ij}=\eta_Q$ is taken for simplicity. Then, the fixed point values of $\lambda_u$, $\lambda_d$ and $\eta_Q$ are estimated as $\lambda_u=\lambda_d \approx 2.42, \ \eta_Q\approx 1.44$.
After $f$ and $\bar f$ are integrated out, the theory becomes stronger but still in the conformal window. The couplings become larger as $\lambda_u = \lambda_d \approx 5.11, \ \eta_{Q}\approx 2.71$ and we obtain a sizable $\lambda_S$ coupling.

\end{document}